\documentclass[aip,apl,amsmath,amssymb,reprint]{revtex4-1}

\usepackage{graphicx}% Include figure files
\usepackage{dcolumn}% Align table columns on decimal point
\usepackage{bm}% bold math
\usepackage[utf8]{inputenc}
\usepackage[T1]{fontenc}
\usepackage{mathptmx}
\usepackage{etoolbox}

\makeatletter
\def\@email#1#2{%
 \endgroup
 \patchcmd{\titleblock@produce}
  {\frontmatter@RRAPformat}
  {\frontmatter@RRAPformat{\produce@RRAP{*#1\href{mailto:#2}{#2}}}\frontmatter@RRAPformat}
  {}{}
}%
\makeatother

\begin{document}

% \preprint{}

\title{LightCode: Compiling LLM Inference for Photonic-Electronic Systems}



\author{Ryan Tomich}
\email{rjtomich@mit.edu}
\author{Zhizhen Zhong}
\author{Dirk Englund}
% \thanks{}
% \altaffiliation{}
\affiliation{Research Laboratory of Electronics, Massachusetts Institute of Technology}


\date{\today}

\begin{abstract}

The growing demand for low-latency, energy-efficient inference in large language models (LLMs) has catalyzed interest in heterogeneous architectures. While GPUs remain dominant, they are poorly suited for integration with emerging domain-specific accelerators like the Photonic Tensor Units (PTUs), which offer low-power, high-throughput linear computation. This motivates hybrid compilation strategies that combine photonic and electronic resources.
We present \textbf{LightCode}, a compiler framework and simulator for mapping LLM inference workloads across hybrid photonic–electronic systems. LightCode introduces the Stacked Graph, an intermediate representation that encodes multiple hardware-specific realizations of each tensor operation. Hardware assignment is formulated as a constrained subgraph selection problem optimized for latency or energy under parametric cost models.
We evaluate LightCode on the prefill stage of GPT-2 and Llama-7B  showing that under our workload and hardware assumptions, 
(i) Photonic hardware reduced energy by up to 50 \% in our simulated workloads at maximum sequence length; 
(ii) multiplexing and assignment strategy yielded latency improvements exceeding $10\times$;
and (iii) Optimizing for latency or energy resulted in distinct hardware mappings in our simulations.
LightCode offers a module, foundational framework and simulator for compiling LLMs to emerging photonic accelerators.
\footnote{The following article has been submitted to \textit{APL Photonics}. After it is published, it will be found at Link.}

% The following article has been submitted to \textit{APL Photonics}. After it is published, it will be found at Link.
\end{abstract}

% \pacs{}% insert suggested PACS numbers in braces on next line

\maketitle %\maketitle must follow title, authors, abstract and \pacs

\section{Introduction}
The development of large language models (LLMs) has led to increased model depth, width, and context length, driving extraordinary  computational demands and making tensor operations a major performance bottleneck \cite{kaplan_scaling_2020}. 
Broader deployment of LLMs in real-world applications have increased the overall volume of inference requests, making latency and energy efficiency critical design considerations \cite{ludvigsen_chatgpts_2023} \cite{brown_language_2020}. 
While training requires enormous compute and memory resources, the inference stage—repeated billions of times in real-world deployments— dominates energy consumption and operational cost \cite{samsi_words_2023, zhong_lightning_2023}. 
These demands drive heavy optimization of GPU-based inference pipelines within common frameworks like PyTorch \cite{ansel_pytorch_2024} and TensorFlow\cite{abadi_tensorflow_2016}.

\begin{figure}[htbp]
    \centering
    \includegraphics[width=1\linewidth]{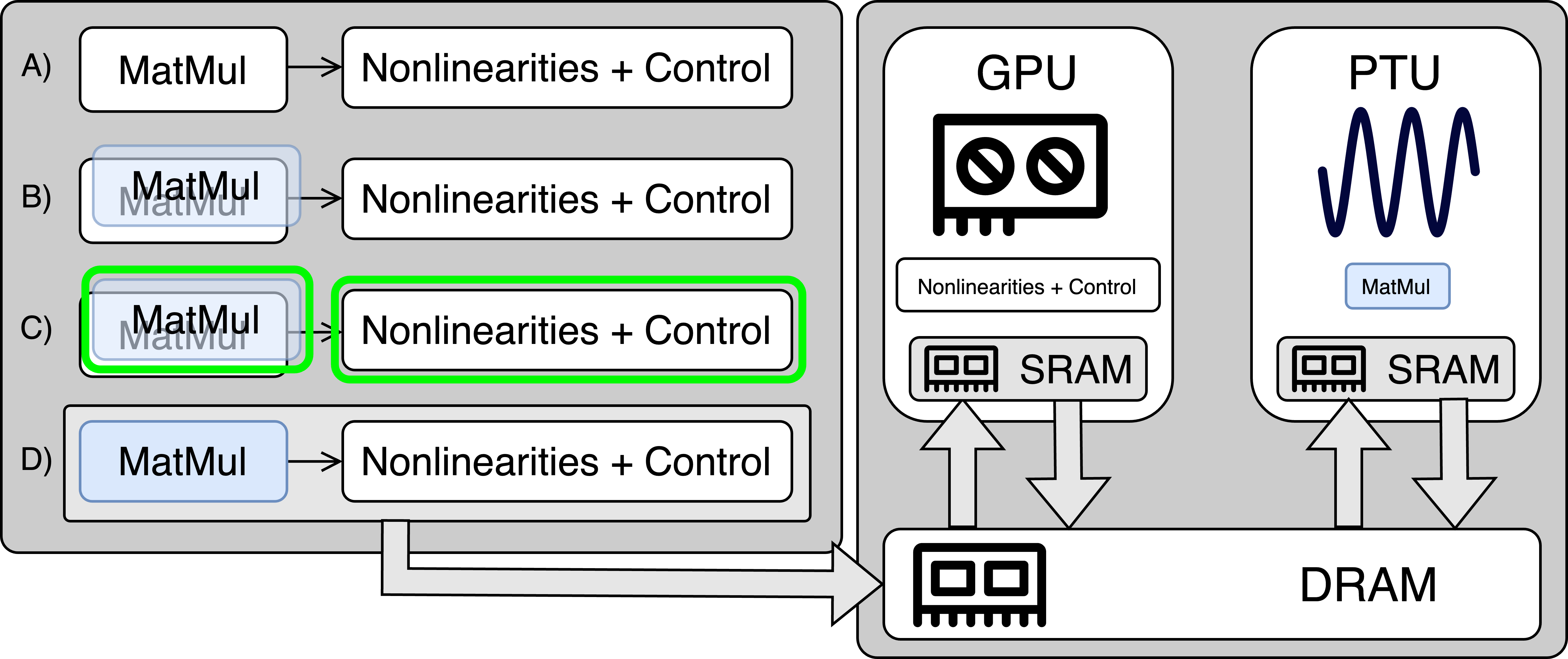}
    \caption{Hybrid execution model with compiler-managed partitioning and scheduling across the GPU and Photonic Tensor Unit (PTU). 
    A) Graph partitioning, B) Stack construction, C) Candidate selection, D) Scheduling.}
    \label{fig:architecture}
\end{figure}

\subsection{Machine Learning Compiler Infrastructure}
ML compilers such as TensorFlow XLA \cite{kaufman_learned_2019}, Apache TVM \cite{chen_tvm_2018}, Glow\cite{rotem_glow_2019}, and MLIR \cite{lattner_mlir_2020} have become foundational tools for optimizing ML workloads. 
These compilers use IRs to apply a range of transformations, including operator fusion, layout optimization, constant folding, and hardware-aware scheduling.
While these compilers target general-purpose hardware, the slowing of Moore’s Law has spurred growing interest in specialized accelerators that promise higher efficiency for domain-specific operations, such as tensor products \cite{jouppi_-datacenter_2017}.

\subsection{Photonic Computing for ML Acceleration}
Large‑scale transformers spend > 95 \% of their time and energy in dense matrix–vector or matrix–matrix multiplies of width 1k–16k.
Recent silicon‑photonic tensor cores can execute those linear layers at <1 pJ per MAC and > 10 TOPS mm$^{-2}$. \cite{feldmann_parallel_2021}
A time/space‑multiplexed $2 \times 2$ coherent X‑bar operated at 20 GBd and 16‑$\lambda$ WDM demonstrates 0.2 pJ MAC$^{-1}$ (0.8 TOPS on 0.04 mm$^{2}$) 
An in‑memory broadcast‑and‑weight microring array reaches 7.3 TOPS mm$^{-2}$ and the same 0.2 pJ MAC$^{-1}$ at 25 Gb s$^{-1}$.\cite{zhou_-memory_2023}
Survey data across >20 tape‑outs place the median integrated MZI mesh at 0.3–3 pJ MAC$^{-1}$ with effective 6‑bit linear precision.\cite{totovic_femtojoule_2020}

However, photonics is not a stand-alone compute platform for LLMs.
A primary limitation is data-converter overhead: a single 6 GS $s^{-1}$ 8-bit ADC/DAC pair dissipates 25–80 fJ per sample, and in full-system models, cross-domain conversion and DRAM traffic often dominate the energy budget—exceeding photonic compute energy by $10\times$. \cite{noauthor_adc_2012}
Additionally, there exists an analog accuracy ceiling: fabrication variability and shot-noise limit the practical dynamic range to approximately 6 ENOB, with experimental studies reporting up to 15 \% accuracy loss when such non-idealities are not properly compensated. \cite{hamerly_asymptotically_2022}
Furthermore, essential nonlinear operations such as activation functions, normalization, attention indexing, and control logic remain electronic, forcing every optical output to be converted back into the electrical domain. 

Consequently,heterogeneous pipelines are currently a widely considered deployment model where photonic cores act as domain‑specific accelerators for the linear sub‑graphs, with GPUs or ASICs handling the surrounding non‑linear and control logic—a design philosophy adopted, for example, by the Albireo light‑offloading prototype. \cite{shastri_photonics_2021}

\subsection{Heterogeneous Compilation and Workload Partitioning}
These heterogeneous pipelines consisting of GPU's and PTU's require coordination via a compiler capable of reasoning across devices with differing Instruction Set Architectures (ISAs), performance characteristics, and energy tradeoffs.
Compiling for heterogeneous hardware has long been a focus in systems such as OpenCL\cite{stone_opencl_2010}, SYCL, and CUDA, which abstract coordination across CPUs, GPUs, and custom accelerators. 
These approaches address challenges such as workload partitioning and task scheduling, but relatively few compilers incorporate PTUs as first-class compilation targets. 

This paper introduces \textbf{LightCode}, a compiler optimization framework and hardware-aware simulator designed to address this challenge. 
LightCode explicitly incorporates photonic compute constraints into the compilation and scheduling pipeline, enabling multi-target LLM inference across photonic and classical hardware. It transforms model execution traces into computation graphs and formulates the hardware mapping problem as a subgraph selection problem. 
LightCode then optimizes a cost function by partitions computation graphs, constructs per-target operation stacks, and selects the lowest-cost hardware mapping under realistic throughput and energy models. 
The main contributions of this paper are:

\begin{itemize}
    \item LightCode Framework: Compiler and simulator for hybrid photonic–electronic LLM inference.
    \item Stacked Graph IR: Novel IR for selecting and mapping ops to optimal hardware.
    \item Evaluation on LLMs: Simulated validation of improved time and energy efficiency on Transformer models.
\end{itemize}

\section{Problem Statement}

We define the hybrid compilation problem for photonic–electronic LLM inference as follows.

\subsection{Workload Setup}
Each model is lowered to a tensor-level dataflow graph \( G = (V, E) \) via the TVM–Relay compiler.  
Each node \( v \in V \) represents an operator \( v_{\text{op}} \) that consumes and produces one or more tensors.  
Each edge \( (u, v) \in E \) corresponds to a data dependency and is weighted by \( s_{u,v} \), the number of bits in the tensor transferred from \( u \) to \( v \).  
The total number of multiply-accumulate (MAC) operations is determined by the input sequence length \( S \). We define these graphs for 2 transformer based LLM's shown in Table~\ref{tab:llm-models}.

\begin{table}[h]
\centering
\caption{LLM workloads. Batch size is one; KV-cache is empty (worst-case prefill throughput).}
\resizebox{\columnwidth}{!}{%
\begin{tabular}{lccccc}
\hline
Model & Layers & $d$ & Params & Stage & $S$ \\
\hline
GPT-2 Small & 12 & 768 & 117M & Prefill & $\{100k \mid k \in \mathbb{N}_0,\ 0 \leq k \leq 10\}$ \\
Llama-7B & 32 & 4096 & 6.7B & Prefill & $\{100k \mid k \in \mathbb{N}_0,\ 0 \leq k \leq 40\}$ \\
\hline
\end{tabular}%
}
\label{tab:llm-models}
\end{table}

\subsection{Hardware Model}
To evaluate alternative mapping strategies, we develop a modular simulator for heterogeneous hardware. Each hardware target in $\mathcal{H} = \{\text{GPU}, \text{PTU}\}$ is modeled as a set of parallel cores with defined \textit{throughput} (instructions per unit time) and \textit{energy cost} (joules per instruction). Data transfers are characterized by \textit{bandwidths} (bits per unit time) and \textit{energy efficiencies} (joules per bit). Hardware parameter’s can be found in Table~\ref{tab:hw-params}.

\begin{table}[htbp]
\caption{\label{tab:hw-params}GPU \cite{noauthor_nvidia_nodate}, PTU, and memory parameters used in simulation. Energy values are reported per operation or bit.}
\centering
\resizebox{\columnwidth}{!}{
  \begin{ruledtabular}
    \begin{tabular}{llc}
      \textbf{Parameter} & \textbf{Symbol} & \textbf{Value} \\
      \hline
      Clock frequency & $f_\text{GPU}$ & 1.98 GHz \cite{} \\
                      & $f_\text{PTU}$ & 9.7 GHz \cite{zhong_lightning_2023} \\
      Energy per MAC  & $E_\text{GPU}$ & 0.07 pJ \cite{zhong_lightning_2023} \\
                      & $E_\text{PTU}$ & 0.04 pJ \cite{zhong_lightning_2023, nahmias_photonic_2020} \\
      Compute cores   & $N_\text{GPU}$ & 144 \\
                      & $N_\text{PTU}$ & 1 \\
      Parallelism     & $W$            & 20\\
      FLOPs per cycle per core & $F^{\text{GPU}}_{cc}$ & 433 \\
      Memory Width      & $b_\text{mem}$ & 32 bits \\
      Memory clock rate & $f_\text{mem}$ & $6 \times 10^9$ Hz (assumed) \\
      DRAM energy     & $E_\text{DRAM}$ & 1 pJ/bit \cite{anderson_optical_2023} \\
      SRAM energy     & $E_\text{SRAM}$ & 0.3 pJ/bit \cite{anderson_optical_2023} \\
      Local memory RW & $E_\text{local}$ & 1 pJ (assumed) \\
      DAC energy      & $E_\text{DAC}$  & 10 pJ/sample \cite{anderson_optical_2023} \\
      ADC energy      & $E_\text{ADC}$  & 3.17 pJ/sample \cite{anderson_optical_2023} \\
      DAC/ADC latency & $\ell_{ij}$     & 10 ns \cite{caragiulo_pietro-caragiulosurvey-dac_2024, murmann_bmurmannadc-survey_2024} \\
    \end{tabular}
  \end{ruledtabular}
}
\end{table}

\subsection{Cost Functions}
Each core type is associated with a set of \textit{translation functions} that map high-level tensor operations to hardware-level instruction sequences. 
This function captures the architectural nuances of each target platform, such as vector widths, specialized compute units (e.g., tensor cores, photonic multipliers), and scheduling constraints.
By applying these mappings to the computational graph, we derive a deterministic cost model for execution time and energy over the entire computation graph.
Inter-core and memory transfers are also explicitly modeled, accounting for latency and energy overheads per bit. 
This includes DRAM reads and writes, HBM accesses, SRAM operations, and transfers involving on-chip local memory. 
Additionally, the framework incorporates the cost of digital-to-analog (DAC) and analog-to-digital (ADC) conversions for photonic data paths.
Given an assignment $\mathbf{x}$ on the computation graph of operations to hardware targets:

Latency mode
$$ c_{\text{lat}}(\mathbf{x})= \sum_{v \in V} t_{h(v)}(v) + \sum_{(u,v) \in E} \tau_{h(u), h(v)}(s_{u,v})$$

Energy mode
$$ c_{\text{eng}}(\mathbf{x})= \sum_{v \in V} e_{h(v)}(v) + \sum_{(u,v) \in E} \epsilon_{h(u), h(v)}(s_{u,v})$$

where
\begin{equation}
\begin{alignedat}{2}
t_{\text{GPU}}(v)       &= \frac{v_{\text{op}}}{F_{cc} \cdot N_\text{GPU}}         &\quad& \text{(GPU compute time)} \\
t_{\text{PTU}}(v)       &= \frac{v_{\text{op}}}{W \cdot N_\text{PTU}}              &     & \text{(PTU compute time)} \\
e_{h}(v)                &= v_{\text{op}} \cdot E_{h}                               &     & \text{(Energy per op on device $h$)} \\
\tau_{ij}(s)            &= \frac{s_{i,j}}{b_\text{mem} \cdot f_\text{mem}}         &     & \text{(Transfer time)} \\
\epsilon_{ij}(s)        &= s_{i,j} \cdot E_\text{local} \cdot E_\text{SRAM}        &     & \text{(Transfer energy)} 
\end{alignedat}
\end{equation}

\section{Methodology}
\subsection{Stacked Graph Intermediate Representation}
\begin{figure}
    \includegraphics[width=0.8\linewidth]{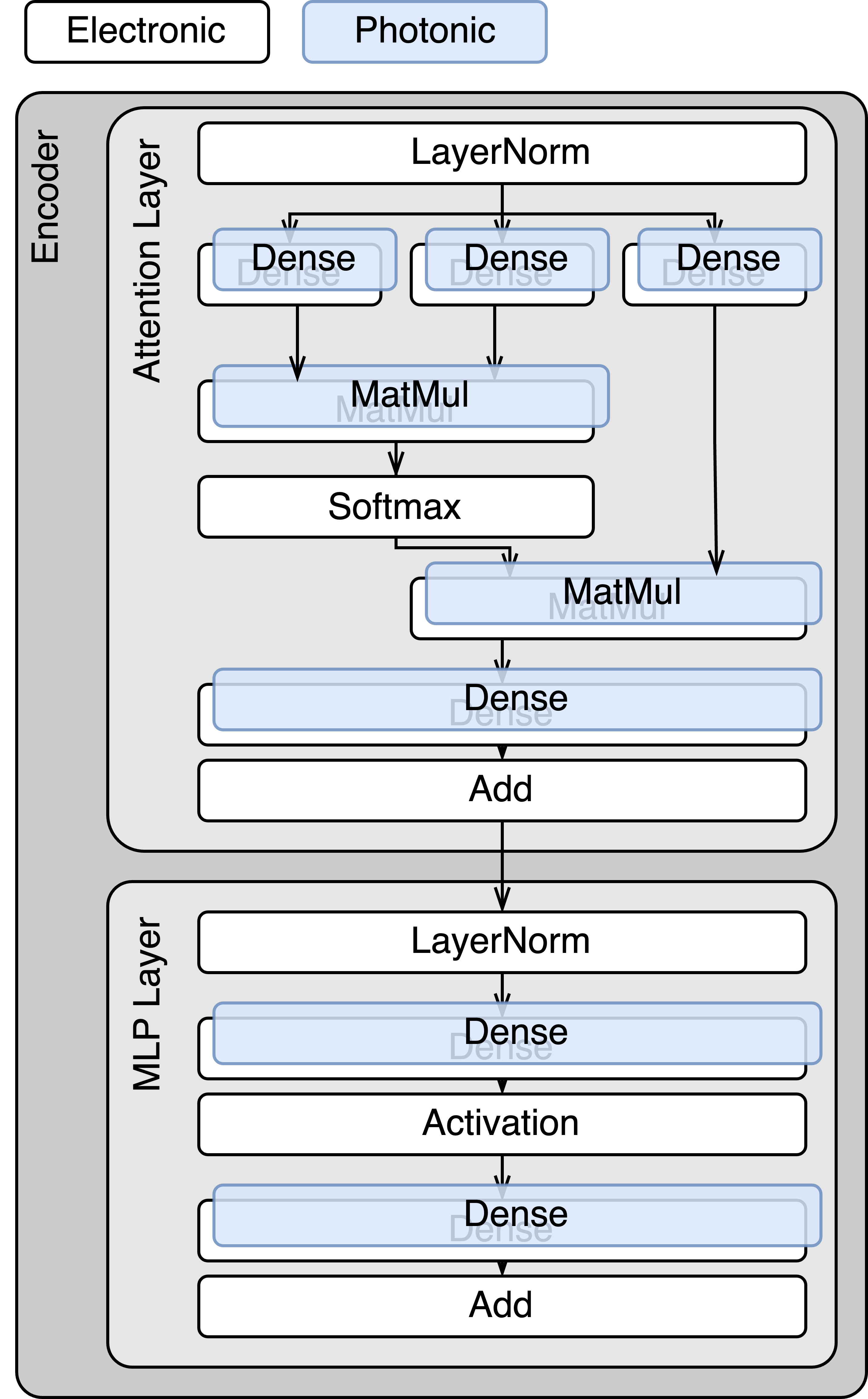}
    \caption{Stacked Graph over a Transformer encoder block, illustrating logical operator grouping across hardware units. 
    This block diagram presents a higher-level abstraction than the Relay IR presented in the Appendix Figure \ref{fig:RelayIR}}
    \label{fig:Stacked_Graph}
\end{figure}

Having established a framework for representing programs as computation graphs over tensor operations, along with a cost model based on hardware-specific arithmetic capabilities, we now introduce a novel IR designed to enable multi-target optimization: the \textbf{Stacked Graph}.
The Stacked Graph extends the traditional computation graph by duplicating each operation node for every supported hardware target. 
Each duplicate represents a hardware-specific implementation of the same logical operation, capturing the unique performance characteristics (e.g., latency or energy) of that backend. 
These alternatives are grouped into \emph{stacks}, where each stack corresponds to a single logical operation and contains multiple candidate implementations—one per target hardware. 
Edges between nodes in adjacent stacks model data dependencies, forming a complete bipartite subgraph for each pair of connected stacks. 
As before, each node carries a weight corresponding to the execution cost of its associated implementation on its target hardware and each edge has a weight that reflects the data transfer cost between pairs of hardware components.
An example of the GPU-PTU stacked graph for the encoding in the transformer architecture is depicted in Figure~\ref{fig:Stacked_Graph}.

Formally, the Stacked Graph transforms a computation graph of operations into a graph of stacks, each encapsulating a set of nodes for mutually exclusive implementation choices. That is, each operation $v$ is assigned to a single node in each stack associated with one hardware target $h \in \mathcal{H}$: 
$$x_{v,h}\in\{0,1\} \text{ where } \sum_{h}x_{v,h}=1.$$
This structure induces a natural optimization problem: select exactly one node from each stack and schedule it such that all inter-stack dependencies are respected and the total cost (execution + transfer) is minimized. This corresponds to a constrained subgraph selection problem over a hypergraph where the global objective is to construct a valid dataflow path with minimum cost.
This optimization problem can be interpreted as a generalization of the \emph{Group Steiner Tree} (GST) problem, in which each group corresponds to a stack and the goal is to select one node per group while minimizing the cost of connecting them. 
While GST is NP-hard in the general case, the repetitive and layered structure of transformer architecture \cite{vaswani_attention_2017} allows for algorithmic simplifications and scalable heuristics. 
In particular, the acyclic and modular topology of transformer computation graphs and the fact that it does not have adjacent linear operations enables graph partitioning strategies to reduce the search space and local greedy choices without sacrificing optimality.

\subsection{Optimization Pipeline}
\paragraph{Computation Graph}
We begin by importing pre-trained model architectures from the Hugging Face model library\cite{wolf_huggingfaces_2020} and lowering them to the TVM Relay IR \cite{roesch_relay_2018, roesch_relay_2019}. 
This produces a computation graph that captures the dataflow dependencies between tensor operations, as defined in the preceding sections.

\paragraph{Partition}
At the tensor level, repeated layers in transformer-based LLMs induce isomorphic subgraphs in the computation graph. These graphs naturally expose articulation nodes—points whose removal increases the number of connected components—providing effective partitioning boundaries. Each component can then be optimized independently, reducing graph size and enabling optimization reuse across isomorphic subgraphs. This greatly improves tractability without compromising optimality, as articulation nodes in transformers under the Rely IR correspond to non-linear operations.

\paragraph{Stack Construction}
For each node in the partitioned graph, we enumerate all viable hardware-specific implementations supported by the available execution backends.
Each node is replaced with a \emph{stack} consisting of an implementation for each feasible hardware target.
Each edge in the original computation graph becomes a set of edges between stacks, connecting all pairs of implementation between stacks. 
This yields the Stacked Graph IR that captures the all posable hardware mappings.

\paragraph{Selection}
We formulate the selection problem as a shortest-path search over the Stacked Graph. 
In the general case, we apply a modified Dijkstra's algorithm to greedily extend shortest paths until a set of complete paths that includes one implementation per stack is identified. 
For transformer models, we exploit the fact that under the Rely IR, linear operations are not adjacent, which allows us to make locally optimal choices independently for each stack with multiple implementations. 
In this case, we restrict the subgraph to the target stack, its parent stacks, and its child stacks. 
% The parent and child stacks only contain one implementation. 
The optimum for the target stack is selected by enumerating all options within that subgraph. 
Global optimality is preserved because linear operations are structurally separated, making each selection independent from the others.
An example of a Relay IR partitioned and scheduled graph is presented in the Appendix under Figure \ref{fig:RelayIR}.

\paragraph{Scheduling}
Once a single implementation has been selected for each stack, the Stacked Graph is flattened into a conventional computation graph by retaining only the chosen nodes and pruning all others. 
The partitioned subgraphs are then merged to reconstruct a unified, executable computation graph. 
We topologically sort the flattened computation graph to respect data dependencies and ensure valid execution ordering. 
Operations are scheduled in that topological order, taking into account backend-specific parallelism constraints and available compute resources. 
The final schedule maps operations to hardware units while honoring dependencies.

\section{Results}

\begin{figure}[htbp]
    \centering
    \includegraphics[width=1\linewidth]{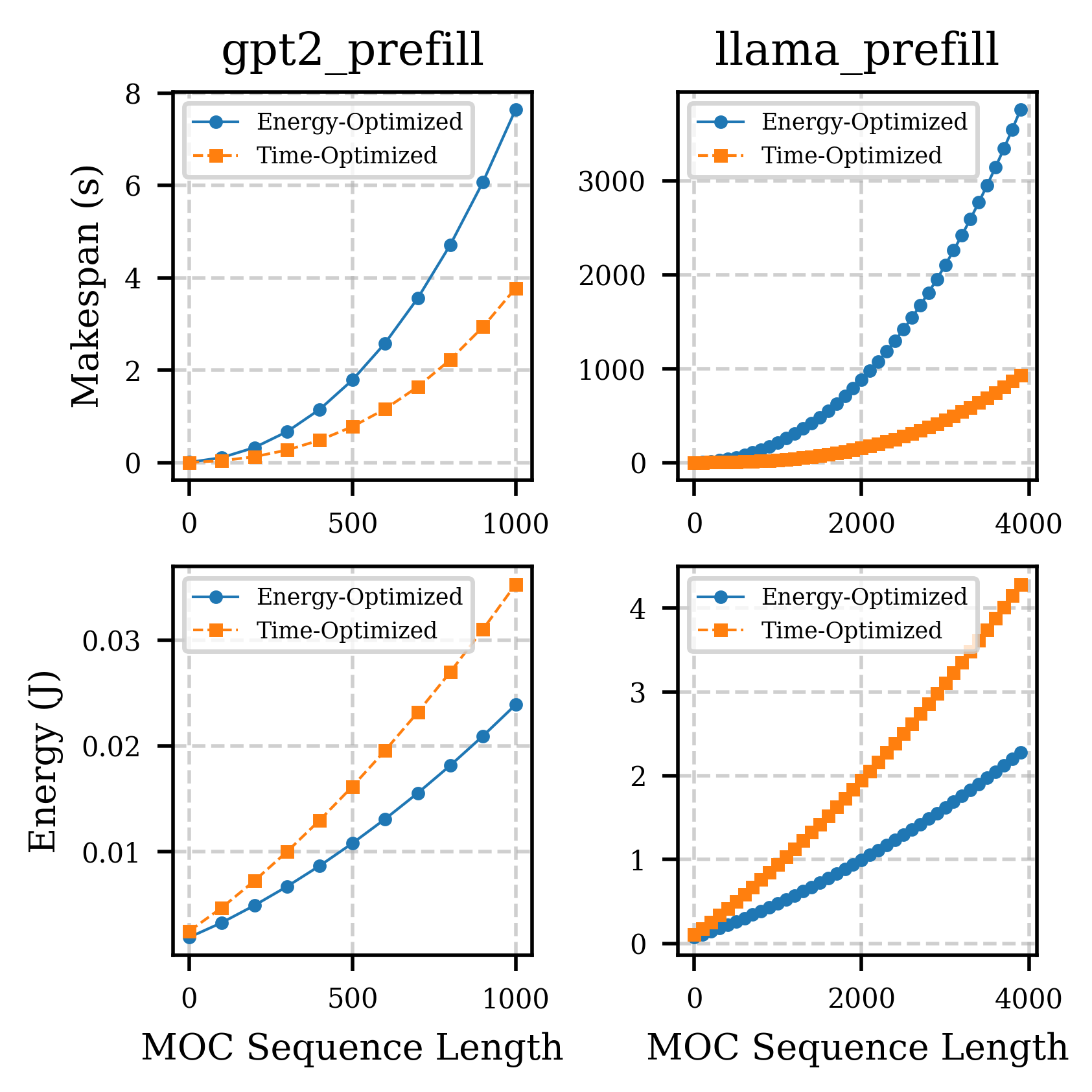}
    \caption{Simulated makespan and energy usage during the prefill stage for Llama-7B and GPT-2 Small (PTU multiplexing = 20).}
    \label{fig:makespan_energy_combined}
\end{figure}

\begin{figure}[htbp]
    \includegraphics[width=0.9\linewidth]{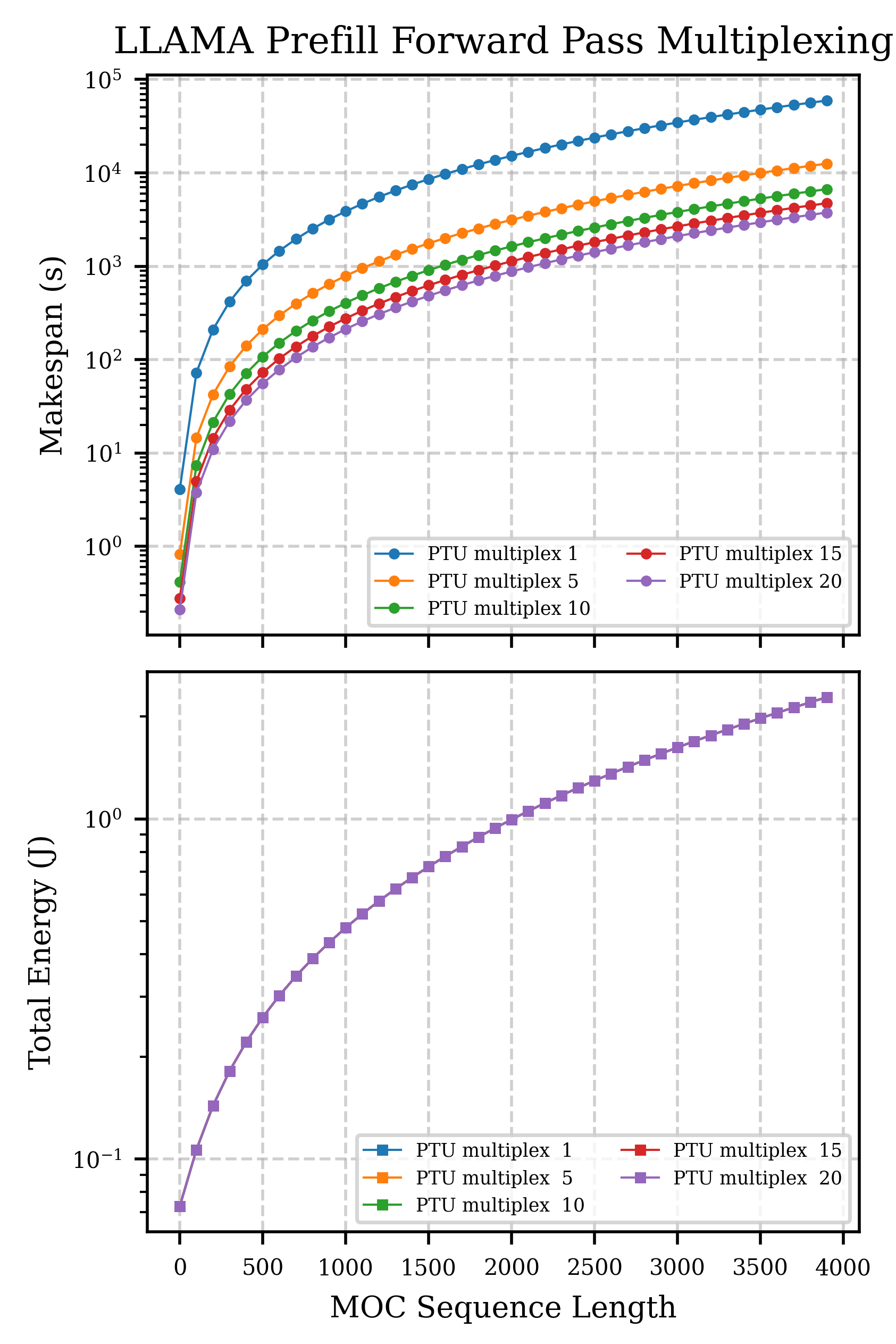}
    \caption{Impact of PTU multiplexing factor on makespan and energy for Llama-7B during the prefill stage.}
    \label{fig:Llama_multiplex}
\end{figure}

\begin{figure}[htbp]
    \centering
    \includegraphics[width=1\linewidth]{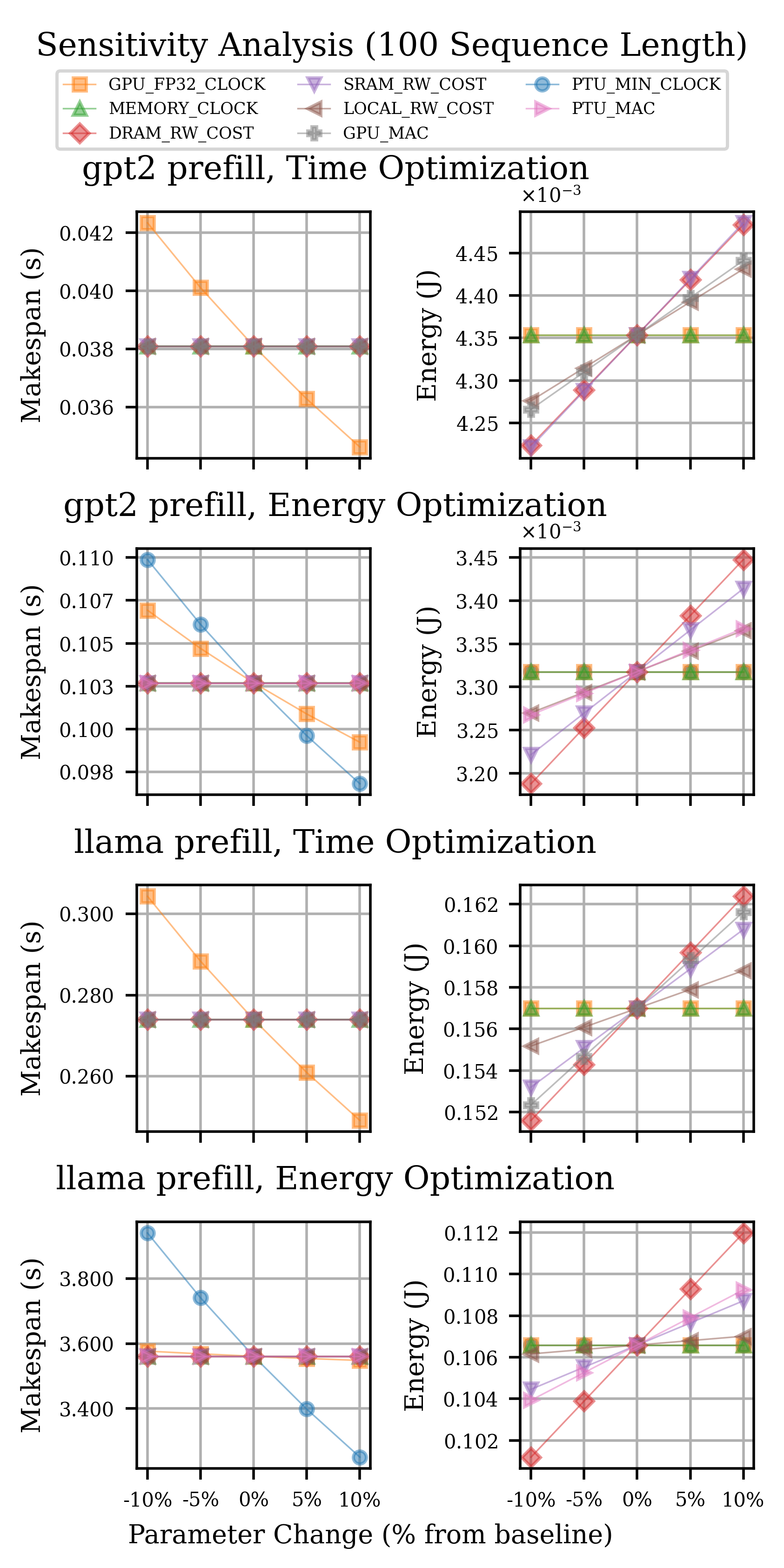}
    \caption{Sensitivity of makespan and energy to hardware parameters for GPT-2 Small and Llama-7B (sequence length = 100) under time- and energy-optimized configurations.}
    \label{fig:sensitivity_analysis}
\end{figure}

\begin{figure}[htbp]
    \includegraphics[width=0.9\linewidth]{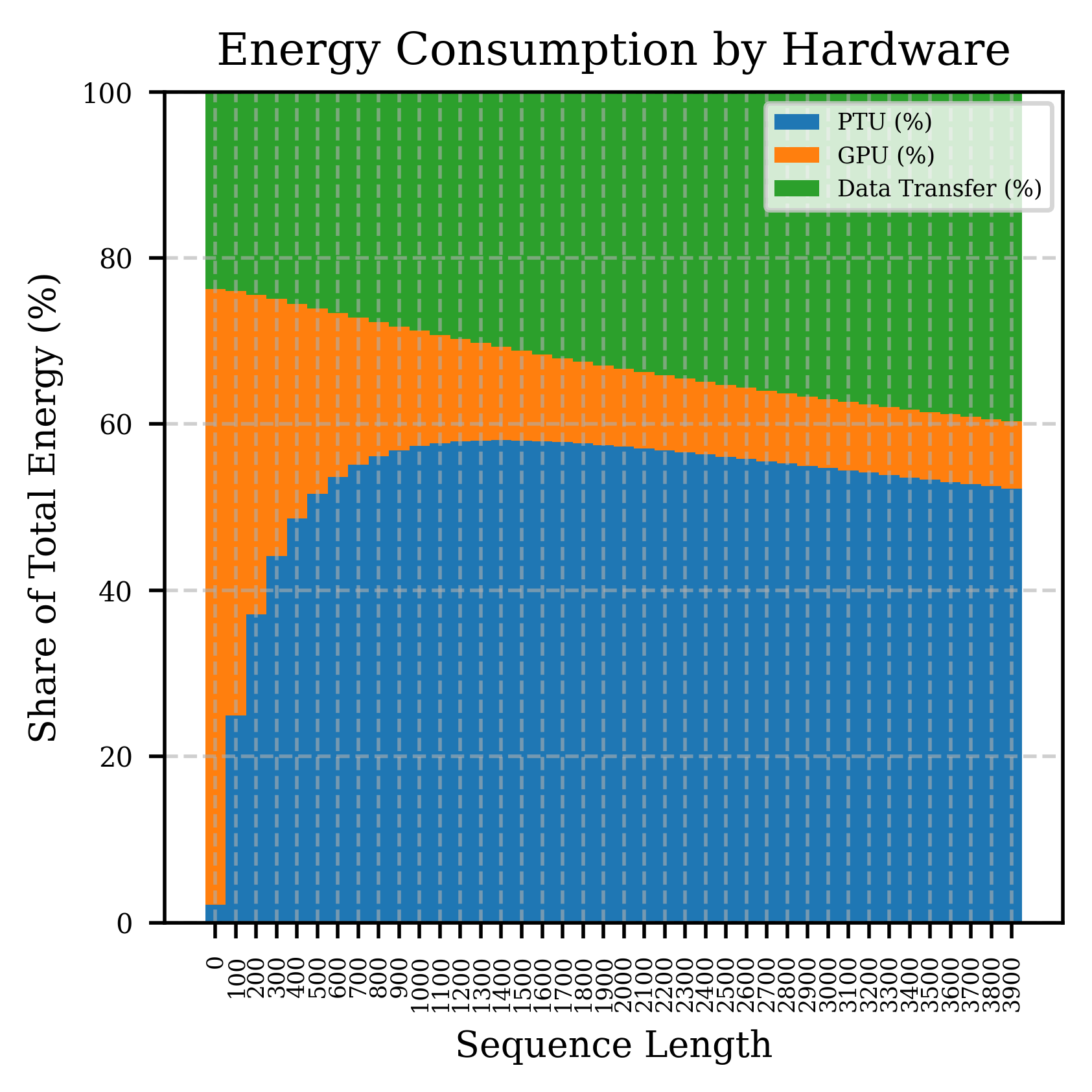}
    \caption{Breakdown of energy consumption by component for Llama-7B across sequence lengths (PTU multiplexing = 20).}
    \label{fig:energy_source_breakdown}
\end{figure}

To evaluate hybrid execution performance, we analyze makespan and energy consumption under LightCode’s hardware mapping strategy (Fig.\ref{fig:makespan_energy_combined}). These results reflect a fixed photonic multiplexing factor of 20. We further explore the impact of multiplexing width on Llama-7B (Fig.\ref{fig:Llama_multiplex}). 
To assess robustness, we conduct a sensitivity analysis over key hardware parameters from Table~\ref{tab:hw-params}, identifying the components most influential to system performance under both latency- and energy-driven objectives(Fig.\ref{fig:sensitivity_analysis}). Finally, Fig.\ref{fig:energy_source_breakdown} disaggregates total energy by subsystem, quantifying the relative contributions of photonic, electronic, and memory elements across input sequence lengths.

\subsection{Comparison to Prior Work}
To validate the accuracy of LightCode’s energy modeling, we compare our results with those from Optical Transformers~\cite{anderson_optical_2023}, which estimates energy based on MAC operation counts and an experimentally obtained measurement for photonic dot-products.

For GPT-2 Small with a sequence length of 200 tokens, our simulation reports $1.0988 \times 10^{11}$ MACs. 
Using a configuration that prioritizes photonic compute, total energy consumption is $0.494 \times 10^{-2}$ J, with $0.102 \times 10^{-2}$ J attributed to photonic cores—comparable to reported figures for GPT-2 Base in Optical Transformers~\cite{anderson_optical_2023}.

For Llama-7B at 400 tokens, the model performs $1.204 \times 10^{13}$ MACs. 
Total energy consumption is $0.2209$ J, with $0.1074$ J attributed to photonic units. 
These values are consistent with the estimates for the GPT-3 6.7B model in Optical Transformers~\cite{anderson_optical_2023, poddar_towards_2025}.

\subsection{Discussion}
Under the stated workload, hardware, and cost model assumptions, Figure~\ref{fig:makespan_energy_combined} show that the optimal hardware placement varies based on the performance metric. 
In our parameter sweep, lowest-latency configurations often keep linear ops on the GPU, while energy-optimized configurations offload them entirely to the PTU.
This simulated result indocates photonic acceleration may not guarantee the optimal trade-off for every objective. Our simulators suggests photonic off-load yields large energy savings, but no latency improvements. The hardware mappings stayed consistent under varying hardware parameters shown in sensitivity analysis in Figure \ref{fig:sensitivity_analysis}.
Under the cost model and hardware assumptions, while increased multiplexing improves performance, Figure~\ref{fig:Llama_multiplex} shows gains start to diminish. We hypothisize that this is due to the assumed constraint that all multiplexed dot products must share a common operand (limiting reuse opportunities), and the observation that PTUs are typically data-bound rather than compute-bound, making memory throughput a critical bottleneck.
Finally, Figure~\ref{fig:energy_source_breakdown} reveals that under our assumptions, photonic dot products and data movement dominate energy consumption as sequence length grows. \cite{zhong_lightning_2023}. 
This underscores the need for improving DAC/ADC efficiency and reducing conversion overheads to fully realize the potential of photonic accelerators.

These findings highlight the nuanced tradeoffs in deploying photonic accelerators for LLM inference. 
While photonic hardware can substantially reduce energy consumption for linear operations, its benefits are constrained by architectural and algorithmic bottlenecks such as memory bandwidth, operand reuse, and conversion overheads. 
The optimal hardware assignment strategy is workload- and objective-dependent: minimizing latency favors conventional GPUs, whereas minimizing energy favors aggressive photonic offloading. 
These findings validate the need for a flexible, workload-aware compiler infrastructure like LightCode capable of exploiting the complementary strengths of hybrid photonic-electronic systems for scalable, efficient LLM inference.

\subsection{Limitations}
\paragraph{Cost Model Simplification}The arithmetic hardware simulator used in LightCode provides a tractable alternative to cycle-accurate simulation (e.g., gem5), but does not model cache behavior, interrupts, or probabilistic execution. 
This may limit accuracy across diverse hardware workloads. 
Additionally, the PTU model does not yet capture noise, drift, or calibration overheads, which may impact performance and energy efficiency in real-world photonic systems. 
The learned cost models in TVM \cite{chen_tvm_2018} were unsuitable due to their focus on ranking rather than physical simulation. 

\paragraph{Hardware Assumptions}
LightCode leverages the structural property of Transformer models in which tensor products are not executed sequentially. 
This make it practical to use a fast, heuristic-based graph search by assuming non-contiguous photonic operations. 
However, for alternative model architectures or future hardware capable of accelerating a broader range of operations, this assumption may not hold, requiring a slower exhaustive search.

\paragraph{Decoupled Selection and Scheduling}
LightCode separates hardware selection from operation scheduling, which can lead to suboptimal pipeline utilization. 
For example, selecting a busy photonic unit while the GPU is idle may increase overall makespan. 
A joint selection-scheduling strategy could improve performance.

\paragraph{Limited Multi-Hardware Parallelism}
Each Relay operation is assigned to a single hardware target, enabling task parallelism but restricting data parallelism. 
While LightCode manually enables photonic matmul tiling, automatic partitioning of operations across multiple hardware types (e.g., splitting a matmul across PTU and GPU) could reduce idle time and improve throughput.

\section{Conclusion}
We present LightCode, a compiler optimization framework and hardware-aware simulator for mapping LLM inference workloads onto hybrid photonic–electronic architectures. As LLMs scale in size and usage, achieving low-latency, energy-efficient inference becomes increasingly critical. Photonic devices offer speed and energy advantages for the linear layers that dominate LLMs, but practical constraints necessitate heterogeneous deployments where photonics function as accelerators alongside GPUs.
To address this challenge, LightCode introduces the Stacked Graph, a novel intermediate representation that encodes hardware-specific operation variants and inter-device transfer costs. Hardware mapping is framed as a constrained subgraph selection problem, enabling principled optimization of execution time and energy across heterogeneous systems.
Simulations on GPT-2 and Llama-7B suggest that selective use of photonic accelerators for dense linear subgraphs yields significant energy savings. However, performance gains appear to depend on workload characteristics, hardware constraints, and optimization objectives—underscoring the need for a flexible, cost-aware compiler.
Looking ahead, LightCode offers a modular platform for exploring compiler strategies at the intersection of machine learning and emerging hardware. Future extensions will integrate tighter scheduling, data-parallelism, and validation on photonic prototypes. By bridging compiler design and heterogeneous hardware, LightCode advances scalable, efficient LLM deployment on next-generation photonic hardware.

% \section{Author Declarations}
% \subsection{Acknowledgements}
% \subsection{Conflict of Interest}
% \subsection{Contributions}
% Ryan Tomich: Conceptualization (lead); Methodology (lead); Software (lead); Validation (lead); Formal analysis (lead); Investigation (lead); Visualization (lead); Writing – original draft (lead); Writing – review and editing (lead).

% Zhizhen Zhong: Supervision (Lead); Methodology (supporting); Writing – review and editing (supporting).

% Dirk Englund: Project administration (lead); Funding acquisition (lead); Supervision (supporting); Writing – review and editing (supporting).

\subsection{Data Availability Statement}
The data and analysis that support the findings of this study are openly available in the GitHub repository at \url{https://github.com/RyanTomich/LightCode}, reference commit \texttt{2b71d8a}.

% \section{}
%\label{}
% \subsection{}
% \subsubsection{}

% If in two-column mode, this environment will change to single-column format so that long equations can be displayed. 
%\begin{widetext}
%$$\mbox{put long equation here}$$
%\end{widetext}

% Figures should be put into the text as floats. 
% Use the graphics or graphicx packages (distributed with LaTeX2e).
% Put the label that you will use with \ref{} command in the braces of the \label{} command.
%
% \begin{figure}
% \end{figure}

% \begin{table}
% \end{tabular}
% \end{table}

%\begin{acknowledgments}
% Put your acknowledgments here.
%\end{acknowledgments}

\section{References}
\bibliographystyle{plain}
\bibliography{lightcode_references}

\appendix
\section*{Appendix: Supplemental Materials}

\subsection{Justification of the Tensor-Level Computation Graph}
In conventional computing systems, an operation is typically defined at the assembly level specified by the ISA.
However, this abstraction is ill-suited for multi-target compilation where various hardware may not aggree on a single ISA.
Moreover, attempting to partition instructions at the abstraction level of an ISA across heterogeneous devices complicates issues such as memory access, register synchronization, and control flow management. 

LightCode adopt a higher-level abstraction aligned with the TVM \cite{chen_tvm_2018} compiler framework: we define an operation as a function that consumes and produces tensor data, as in the TVM Relay IR \cite{roesch_relay_2018, roesch_relay_2019}.
Under this model, a computation graph represents the dataflow dependencies between tensor operations, such as matrix multiplications, element-wise activations, and normalization layers. 
Each node in the computation graph is parameterized by input tensor shapes, the operation performed, and the resulting output shapes. 
These parameters allow the compiler to reason about hardware feasibility, instruction count, data movement costs, and parallelization opportunities. 
This tensor-centric representation provides a more flexible and hardware-agnostic foundation for compiler optimizations. 
It enables operations to be scheduled across hardware targets while maintaining functional correctness. 
This graph structure defines the search space for optimizations, particularly for mapping key primitives—such as tensor products—to specialized accelerators.

\begin{figure}[htbp]
    \centering
    \includegraphics[width=1\linewidth]{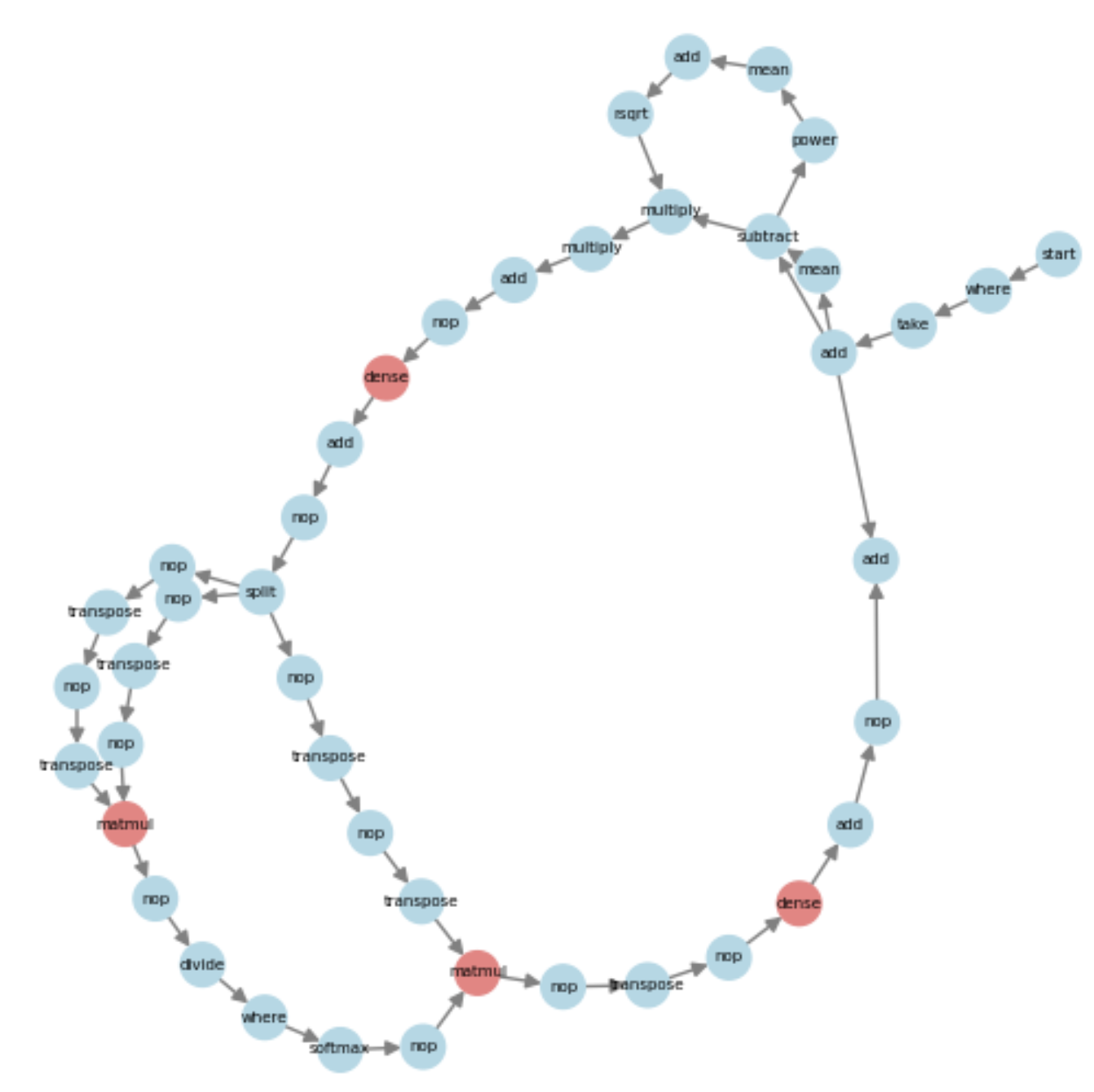}
    \caption{Example partition 1 of the GPT-2 Small Transformer in Relay IR. A subset of the computation graph is shown after partitioning. Highlighted nodes are maped to PTU. }
    \label{fig:RelayIR2}
\end{figure}

\begin{figure}[htbp]
    \centering
    \includegraphics[width=1\linewidth]{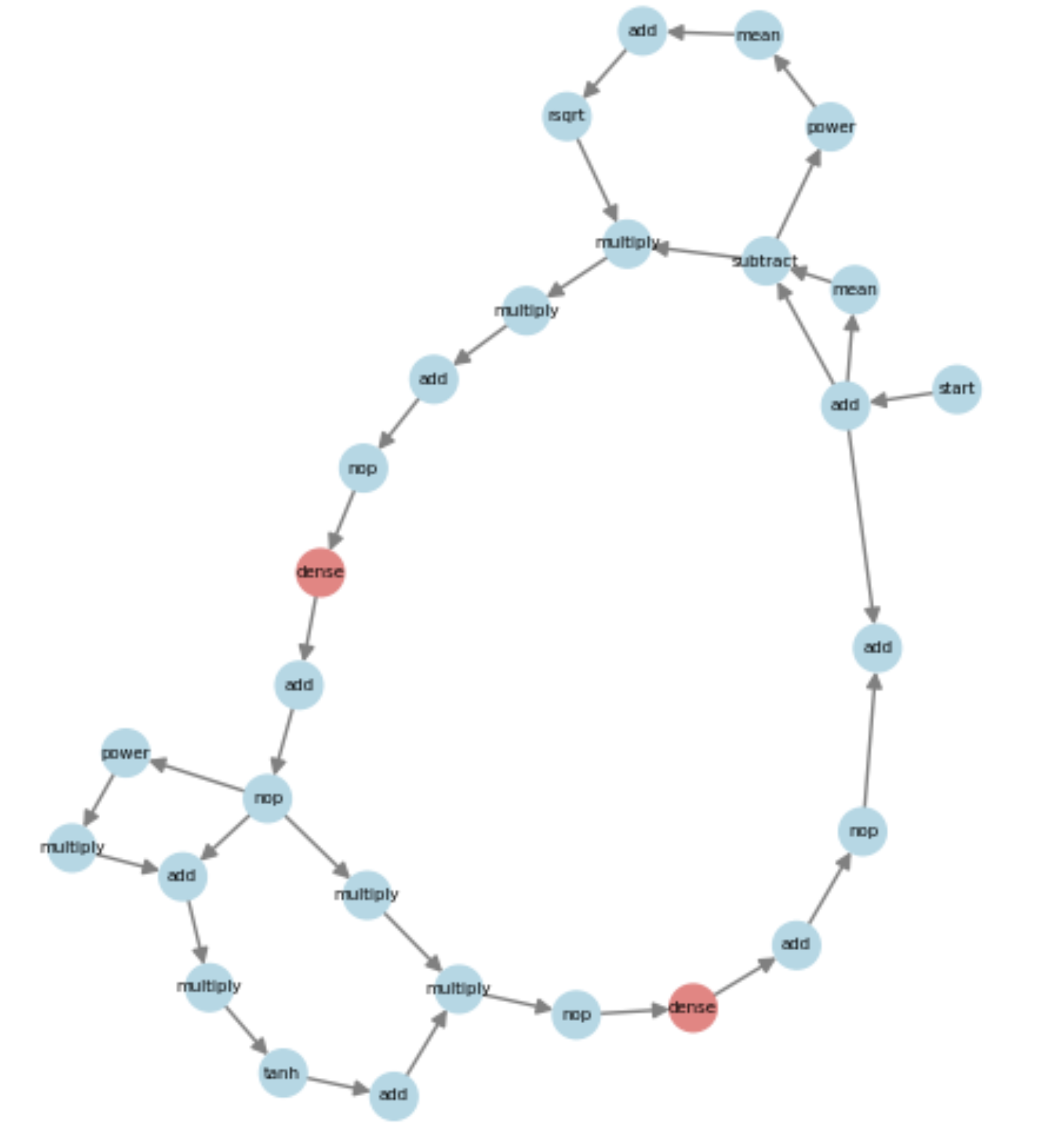}
    \caption{Example partition 2 of the GPT-2 Small Transformer in Relay IR. A subset of the computation graph is shown after partitioning. Highlighted nodes are maped to PTU. }
    \label{fig:RelayIR}
\end{figure}

\end{document}